\documentclass[conference]{IEEEtran}
\IEEEoverridecommandlockouts
\usepackage{cite}
\usepackage{amsmath,amssymb,amsfonts}
\usepackage{algorithmic}
\usepackage{graphicx}
\usepackage{textcomp}
\usepackage{xcolor}

\usepackage{textcomp}
\usepackage{color, colortbl}
\usepackage{array, makecell}
\definecolor{Gray}{gray}{0.9}
\definecolor{LightCyan}{rgb}{0.88,1,1}
\definecolor{DarkGray}{rgb}{0.66, 0.66, 0.66}
\definecolor{AshGrey}{rgb}{0.6,0.73,0.89}
\usepackage{easyReview}

\def\BibTeX{{\rm B\kern-.05em{\sc i\kern-.025em b}\kern-.08em
    T\kern-.1667em\lower.7ex\hbox{E}\kern-.125emX}}

\begin{document}

\title{\emph{RDMSim}: An Exemplar for Evaluation and Comparison of Decision-Making Techniques 
for Self-Adaptation
}

\author{
\IEEEauthorblockN{Huma Samin, Luis H. Garcia Paucar, Nelly Bencomo \IEEEauthorrefmark{1}\\
Cesar M. Carranza Hurtado\IEEEauthorrefmark{2}, Erik M. Fredericks\IEEEauthorrefmark{3}}
\IEEEauthorblockA{\IEEEauthorrefmark{1}SEA, Aston University, Birmingham, UK \emph{\{h.samin, garcial2, n.bencomo\}@aston.ac.uk}}
\IEEEauthorblockA{\IEEEauthorrefmark{2}Universidad Pontificia Católica del Perú, Lima, Perú \emph{cesarmiguelch@hotmail.com}}
\IEEEauthorblockA{\IEEEauthorrefmark{3}Grand Valley State University, Michigan, USA \emph{frederer@gvsu.edu}}
}

\newcommand*{\nb}[1]{{\color{magenta}(NELLY: #1)}}
\newcommand*{\hs}[1]{{\color{orange}(HUMA: #1)}}
\newcommand*{\lhgp}[1]{{\color{blue}(LUIS: #1)}}
\newcommand*{\todolist}[1]{{\color{red}(TODO: #1)}}

\maketitle

\begin{abstract}
Decision-making for self-adaptation approaches need to address different challenges, including the quantification of the uncertainty of events that cannot be foreseen in advance and their effects, and dealing with conflicting objectives that inherently involve multi-objective decision making (e.g., avoiding costs vs. providing reliable service). 
To enable researchers to evaluate and compare decision-making techniques for self-adaptation, we present the \textit{RDMSim} exemplar. 
\textit{RDMSim} enables researchers to evaluate and compare techniques for decision-making under environmental uncertainty that support self-adaptation. The focus of the exemplar is on the domain problem related to Remote Data Mirroring, which gives opportunity to face the challenges described above. 
\textit{RDMSim} provides probe and effector components for easy integration with external adaptation managers, which are associated with decision-making techniques and based on the MAPE-K loop. 
Specifically, the paper presents 
(i) \textit{RDMSim}, a simulator for real-world experimentation, 
(ii) a set of realistic simulation scenarios that can be used for experimentation and comparison purposes,
(iii) data for the sake of comparison. 

\end{abstract}


\section{Introduction}

Remote Data Mirroring (RDM) is a disaster recovery technique used to protect data by storing multiple copies (i.e. replicas) on physically remote servers (i.e. mirrors) \cite{ji2003seneca,Keeton04}. The RDM system tolerates failures by requesting or rebuilding the lost or damaged data samples from another active mirror to facilitate data recovery. Hence, the RDM helps in maintaining data availability and preventing data loss.  Furthermore, to ensure that distributed data is not lost or corrupted, the RDM is required to perform the replication and distribution of data in an efficient and reliable way.

Considerable research efforts have targeted the domain of Remote Data Mirroring \cite{Ramirez2012b,Fredericks2015,cure_chapter_2015,Paucar2020,samin2020priority,BowersFredericksCheng2018}. 
However, the RDM applications  are very costly to implement as the equipment used to install such applications is expensive. To the best of our knowledge, there is no exemplar available to support research based on the RDM paradigm.


In this paper, we present \textit{RDMSim}, an exemplar that simulates a 
Remote Data Mirroring environment. The goal of the \textit{RDMSim} is to offer researchers a 
RDM environment to test and compare their decision-making techniques \cite{de2013software} against other techniques. 
Other exemplars exist however, they focus on other domains and aspects, such as cloud environments \cite{barna2015hogna},  cyber-physical systems \cite{kit2015architecture},  traffic management system \cite{schmid2017model}, client-server systems \cite{cheng2009evaluating} and IoT-based systems\cite{IftikharRBW017}. In comparison to \cite{barna2015hogna}, that deals mainly with the functionality of cloud environments such as workload management using the addition and removal of virtual machines, \textit{RDMSim} focuses mainly on the simulation of Remote Mirroring process. 

The \textit{RDMSim} exemplar presented here is implemented in Java, keeping in view the operational model presented in \cite{Keeton04,ji2003seneca}. It simulates the RDM presenting a fully connected network of mirrors. The simulator offers the flexibility of changing the number of mirrors to create a customized RDM network according to the experiment's requirements. The focus is on the application of self-adaptive realization strategies in the form of the topologies of Minimum Spanning Tree (MST) and Redundant Topology (RT). The application of these topologies have an impact on the different network parameters such as bandwidth consumption and active network links affecting the quality objectives such as the minimization of operational costs and the maximization of the reliability of the network. A trade-off of such impacts has to be taken into account as part of the decision making\cite{elahi2011requirements,saadatmand_fuzzy_2015,ramirez2009evolving,sawyer_requirements-aware_2010, Goldsby2008}. The topological impacts have been defined based on the expert knowledge presented in \cite{Paucar2020}. Additionally, we provide an implementation of different scenarios that define possible different uncertain environmental contexts for the RDM\cite{esfahani_taming_2011}. A Python version is also publicly available. Researchers can use these scenarios to test their specific decision-making techniques based on Reinforcement Learning \cite{samin2021priority}, Multi-Criteria Decision Analysis \cite{triantaphyllou2000multi} and Evolutionary Computation \cite{Ramirez2012b} among others. Researchers can also design their own scenarios by modifying the different parameter ranges.

The paper is organized as follows: Section 2 presents the operational model of a RDM. 
In Section 3, we present the architecture of the \textit{RDMSim} exemplar. Section 4 provides a description of the different scenarios for the experiments that can be executed by the \textit{RDMSim}. In Section 5, an example of how to execute experiments using \textit{RDMSim} is provided, which is followed by Conclusion in Section 6.

\section{Remote Data Mirroring} 
The RDM application is composed of data servers and network links~\cite{ji2003seneca,Keeton04}. It must replicate and distribute data in an efficient manner by minimizing consumed bandwidth and providing assurance that distributed data is neither lost or corrupted \cite{ji2003seneca}.  
The RDM application must achieve functional objectives such as \textit{construct a connected network} and \textit{distribute data}. These functional objectives can be achieved through alternative realization strategies represented by two different topologies:  \textit{Minimum Spanning Tree} (MST) and \textit{Redundant Topology} (RT). An MST Topology uses the least possible number of network links to transmit data among different remote servers. Contrarily, an RT topology uses simultaneously,  several redundant network links paths to transmit information among remote servers. 

The implementation of the RDM considered in this paper should also satisfy the following three quality objectives: \textit{Maximization of Reliability} (MR), \textit{Maximization of Performance} (MP) and \textit{Minimization of Cost} (MC).
The levels of satisfaction associated with reliability, performance and cost of the RDM are determined according to the trade-offs based on:
\begin{itemize}
    \item A RT Topology \textit{offers higher levels of reliability} than an MST topology. However, the cost of maintaining an RT topology may be prohibitive in some contexts, given the additional cost of bandwidth consumption required.
    \item Conversely, a MST Topology \textit{offers higher levels of performance} with \textit{lower levels of cost} than an RT topology. However, the reliability of the system can be impacted in a negative way when an MST Topology is used. 
\end{itemize}

Based on the above, we have designed the \textit{RDMSim} exemplar. Next, we present the architecture for \textit{RDMSim}.

\section{Architecture}

\begin{figure}[h!]

\centering
\includegraphics[width=.45\textwidth,keepaspectratio]{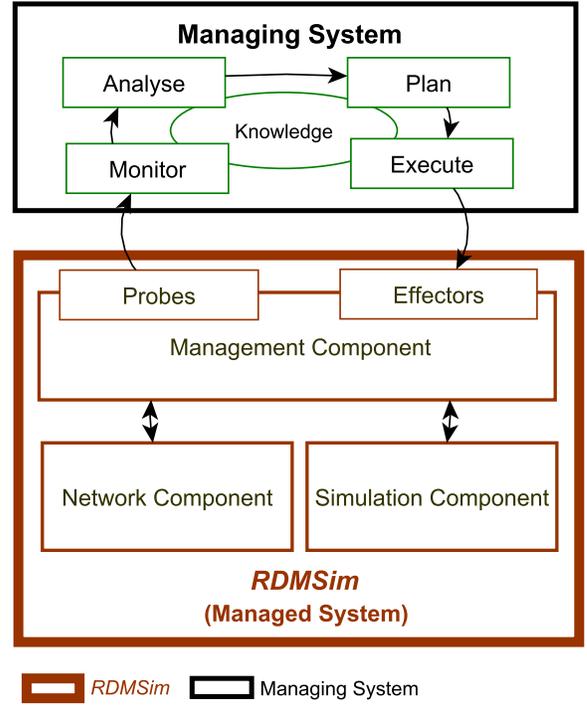}
\vspace{-2mm}
\caption{RDMSim Architecture}
\label{figRDMSimarchitecture}

\end{figure}

The \textit{RDMSim} exemplar has been developed to facilitate the implementation of a two-layered architecture for a self-adaptive RDM, as shown in Fig. \ref{figRDMSimarchitecture}. The architecture structures a Managing System (based on feedback loop \cite{Kephart2003,Brun2009}) on top of the Managed System (the \textit{RDMSim}). We next describe each layer.

\subsection{\textbf{Managing System}}
The Managing System, at the upper layer, is responsible for providing the self-adaptive decision-making logic. A feedback loop is implemented to monitor the environment and managed system, adapting the latter when necessary. 
 The feedback loop consists of  Monitor-Analyse-Plan-Execute over a Knowledge base K (MAPE-K) \cite{Kephart2003}. The MAPE-K loop is considered an architectural blueprint for self-adaptive systems and is used to perform adaptation decisions on the Managed System (i.e. \textit{RDMSim} in our case). When using the \textit{RDMSim} exemplar, researchers will provide their own decision-making techniques to serve as a Managing System. The Managing System can be based on different techniques such as Multi-Criteria Decision-Making~\cite{triantaphyllou2000multi}, Reinforcement Learning~\cite{samin2020priority}, and Evolutionary Computation~\cite{Ramirez2012b,BowersFredericksCheng2018}, etc.

 \subsection{\textbf{Managed System} }
\textit{RDMSim} represents the Managed System and provides probes and effectors that can be used by the Managing System to interact with the simulator. Probes are used to monitor information (M in MAPE) whereas the effectors are used to execute the adaptation decisions (E in MAPE) on the Managed System.

Next, we present the architecture of the Managed System  implemented as Java Packages for the \textit{RDMSim} software. 
The components in the architecture for \textit{RDMSim}, presented in Fig.\ref{figRDMSimarchitecture}, are as follows:


\subsubsection{\textbf{Management Component}} which acts as a bridge between the Managing System and other internal components of the \textit{RDMSim}. It provides an implementation of probes and effectors to be used by the Managing System.  
The functions provided by the probes and effectors are used to both monitor the status of the RDM (i.e. cost, reliability and performance) and also change the network topology and different network parameters according to the decision made as described in Table \ref{tab2} and \ref{tab3} respectively.

\begin{table}
\caption{Probe Functions }\label{tab2}
\vspace{-4mm}
\begin{center}
\centering\renewcommand\cellalign{lc}
\fontsize{6}{8}\selectfont
\begin{tabular}{|l|l|l|}
\hline
\rowcolor{AshGrey}
\textbf{Function} & \textbf{Description} \\
\hline
Topology getCurrentTopology() &\makecell{Returns the current topology\\ for the network.}  \\
\hline
 int getBandwidthConsumption() & \makecell{Returns the bandwidth consumption \\of the network.} \\
\hline
 int getActiveLinks()& \makecell{Returns the number of active links.} \\
\hline
int getTimeToWrite()& \makecell{Returns the time to write data \\for the network.}  \\
\hline
Monitorables getMonitorables()&\makecell{ Returns the values for all the \\monitorable metrics.}\\
\hline

\end{tabular}
\end{center}

\end{table}

\begin{table}
\caption{Effector Functions }\label{tab3}
\centering\renewcommand\cellalign{lc}
\fontsize{6}{8}\selectfont
\vspace{-2mm}
\begin{tabular}{|l|l|l|}
\hline
\rowcolor{AshGrey}
\textbf{Function} & \textbf{Description} \\
\hline
void setNetworkTopology(int timestep,Topology selectedtopology)& \makecell{To set the network\\ topology at a \\particular timestep. } \\
\hline
void setActiveLinks(int active\_links)& \makecell{to set the number of \\active links for \\the network.}    \\
\hline
void setTimeToWrite(double time\_to\_write)& \makecell{To set the time to write \\data for the \\network.} \\
\hline
 void setBandwidthConsumption(double bandwidth\_consumption)& \makecell{To set bandwidth \\consumption for the\\ network.}\\
\hline
 void setCurrentTopology(Topology current\_topology)&\makecell{ To set topology for the \\network.}\\
\hline

\end{tabular}

\end{table}

\subsubsection{\textbf{Network Component}} which provides an implementation of the main physical elements of the RDM. These elements include  the number of mirrors (i.e. servers) and the network links that represent a fully connected network of mirrors. As an example, for 25 mirrors, a network of 300 links will be created.  The users of \textit{RDMSim} can change the number of mirrors to create a custom RDM network for their experiments. 
The Network Component also provides an implementation of the monitorables and topologies for the network.  
Specifically, in the \textit{RDMSim}, we provide an implementation of three monitorables:

\textit{\textbf{Mon1--} \textbf{Active Network Links:}} provides the current active network links to measure the reliability of the RDM. The RDM will provide a higher level of reliability with a larger number of active links.

\textit{\textbf{Mon2--} \textbf{Bandwidth Consumption:}} provides the current bandwidth consumption  to measure the operational cost for the RDM in terms of inter-site network traffic. Operational costs will be increased for the RDM with a higher amount of bandwidth consumed. Bandwidth Consumption is measured in GigaBytes per second.  

\textit{\textbf{Mon3--} \textbf{Time to Write Data to mirrors:}}  measures the performance of the network in terms of writing time to maintain multiple copies of data on each remote site. A big writing time leads to reduction of performance of the RDM. Time to Write Data is measured in milliseconds.

For the communication between the mirrors, we consider synchronous mirroring \cite{cure_chapter_2015,Keeton04}.  During synchronous mirroring, sequential writing is performed to prevent data loss \cite{cure_chapter_2015}. In sequential writing, the primary mirror (i.e. the sender) waits for an acknowledgement (known as a \textit{handshake}) regarding the receipt and writing of data from the secondary mirror (i.e. the receiver). This process is performed for each active link on the communication path between the mirrors. 
Therefore, the time to write data is computed as \textit{Total Writing Time= (\begin{math}\alpha\end{math}* number of active links) * Time to Write Data Unit}\footnote{To implement realistic impacts, we vary the time between 10 to 20 milliseconds}. Here, \begin{math} \alpha \end{math} represents a fraction of active links to constitute the communication path between mirrors. \begin{math} \alpha \end{math} can have a value of greater than zero and less than and equal to one.  For our experiments, we have set \begin{math} \alpha = 1\end{math}.

Similarly, the bandwidth consumption is also dependent on the number of active links. More active links imply more data transmission, which leads to a higher bandwidth consumption \cite{cure_chapter_2015}. Hence, we compute the Bandwidth Consumption as \textit{Total Bandwidth Consumed=(\begin{math}\alpha * \end{math}number of active links) * Bandwidth per link}\footnote{To implement realistic impacts we vary the Bandwidth per link between 20 to 30 GBps}.

\subsubsection{\textbf{Simulation Component}}
which includes the implementation of the uncertainty scenarios\cite{esfahani_taming_2011,giese_living_2014} that represent the different dynamic environmental conditions that the RDM can face, and which will be simulated. It allows the setting of the simulation properties, such as the number of simulation runs and the chosen uncertainty scenario(s) to be executed by the \textit{RDMSim}.

\begin{figure*}[h!]
\centering

\includegraphics[width=\textwidth,height=8cm]{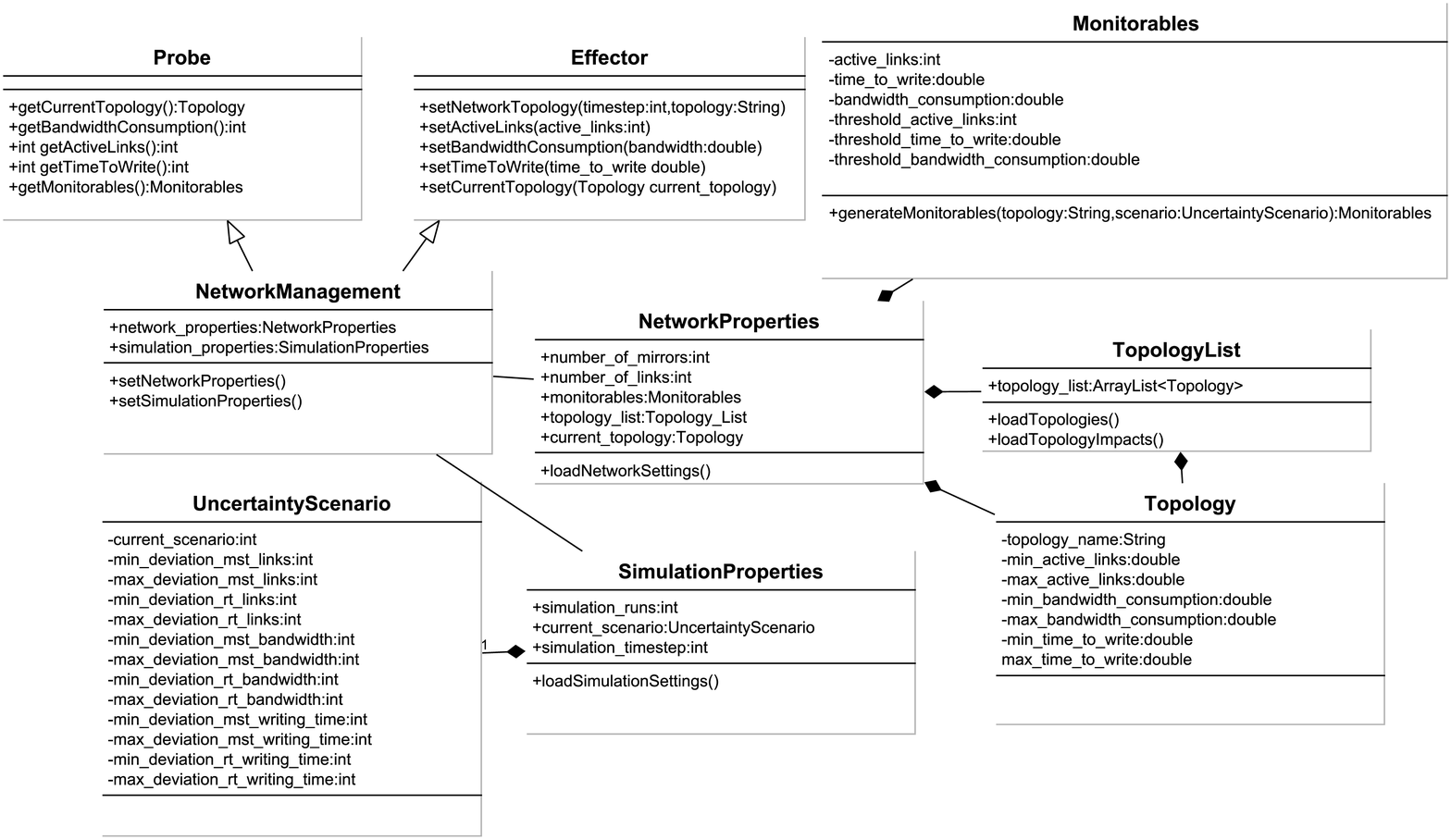}
\caption{RDMSim Class Diagram}
\label{figRDMClassdiagram}

\end{figure*}

A partial class diagram representing the elements of the Management Component, Network Component and Simulation Component is shown in Fig. \ref{figRDMClassdiagram}. The \textit{NetworkManagement} class along with the \textit{Probe} and \textit{Effector} interfaces  provides an implementation of the Management Component. The classes \textit{NetworkProperties}, \textit{Monitorables}, \textit{Topology} and \textit{TopologyList} are part of the Network Component and  provide an implementation of the corresponding features of the RDM. The \textit{SimulationProperties} and \textit{UncertaintyScenario} classes are part of the Simulation Component, and are used to implement the functionalities related to the simulations to be executed.

\section{Scenarios}


Six different scenarios were defined to be used in  simulations of the RDM. These scenarios have been designed to simulate different archetypal  real  situations,  which can cause  deterioration of  satisfaction  of  the quality objectives of the system in  relation  to  a scenario with stable conditions.



The main goal of the scenarios depicted below,  is to evaluate how decision-making techniques and algorithms react under uncertain situations, specially different from the stable conditions.  
Next, a description for each scenario is presented.\\

 
	    \textbf{Default scenario S$_{0}$: } For the sake of comparison between techniques, a default scenario is provided that represents an environment envisioned by the requirements experts\cite{Fredericks2015,Paucar2020}. For the \textit{RDMSim}, the following thresholds  for the levels of satisfaction associated with reliability, performance and cost are suggested: bandwidth consumption should be on average less than or equal to 40\%. Similarly, the time to write data should be on average less than or equal to 45\%. On the other hand, the number of active links should be on average greater than or equal to 35\% of the total number of links.  
	    The initial topology being used is MST topology.\\ 
		
		\textbf{Scenario S$_{1}$ - Unexpected Packet Loss during MST: }
		The initial topology being used is MST Topology. A period of consecutive and unexpected data packet loss during the execution of the MST Topology generates a reduction on the reliability of the system. 
		Data packet loss represents link failures in the RDM system, which may be caused, for example, by problems with the equipment (e.g. failures in a switch or router or power failures \cite{ji2003seneca}). \\

		\textbf{Scenario S$_{2}$ - Unexpected Packet Loss during RT: }
		The initial topology being used is RT Topology.	Unexpected data packet loss during the execution of the RT Topology, are generating an unusual rate of data forwarding, which would increase the bandwidth consumption (i.e. cost), and would reduce the system's performance. 	
		As said before, in the RDM, the cost for inter-site links communication is a function of the data sent over them. Therefore, a \textit{Redundant Topology} (RT), which involves a bigger number of inter-site network links than a \textit{Minimum Spanning Tree Topology} (MST), is more expensive. Cost increases as the number of active links increases and a reduction on the system's  performance\footnote{The performance in these systems is measured as the total time to perform the write of data, which is the sum of the response times of the writes of each copy of data on each remote site \cite{ji2003seneca}.} could also be expected. \\

		\textbf{ Scenario S$_{3}$}. Simultaneous occurrence of the scenario $S_{1}$ and $S_{2}$. The current topology is randomly generated.\\

	\textbf{Scenario S$_{4}$ - MST topology execution failures: } The topology being used is MST Topology. Involves the behaviour presented in the scenario S$_{1}$. Additionally, during the execution of the MST topology, an increment in bandwidth consumption (MC) and the reduction of the system's  performance (MP) is also produced due to synchronous mirroring.\\ 

		\textbf{Scenario S$_{5}$ - RT Topology execution failures}. The topology being used is RT Topology. Involves the behaviour presented in the scenario S$_{2}$. Additionally, the RT topology is also producing a reduction on the reliability of the system (MR) due to failures in the equipment such as routers and switches.\\
		
		\textbf{Scenario S$_{6}$ - Significant site failure}. The current topology is randomly generated. This scenario involves the simultaneous occurrence of the scenarios $S_{4}$ and $S_{5}$. 
    It is related to a significant site failure~\cite{ji2003seneca,Keeton04}, where both, repeated and multiple concurrent failures are expected~\cite{ji2003seneca}  as in the scenarios S$_{4}$  and S$_{5}$ but all at the same time. A full-scale site failure may be caused by a power outage affecting all the buildings on different campuses, an earthquake or flood affecting buildings within several metropolitan areas. Under this scenario, the worst-case data loss \cite{Keeton04} may occur in different sites (RDM nodes), i.e. a site can be destroyed or inoperative before the full backup of information is shipped offsite. Site failure disasters are usually modelled with a failure rate of once per year \cite{Keeton04}. 

\section{Experiments}

In this section, we provide a simple example to describe the steps to develop a custom adaptation logic for performing experiments using the \textit{RDMSim}. We also demonstrate the execution of different uncertainty scenarios using the custom adaptation logic.

The steps for developement of custom adaptation logic are as follows: 

\subsection*{\textbf{Step: 1 Download the RDMSim Exemplar}} Download the \textit{RDMSim} package from the \textit{RDMSimExemplar} repository
S\footnote{https://doi.org/10.5281/zenodo.4613152}
and install the required libraries.

\subsection*{\textbf{Step: 2 Design an Adaptation  Solution}}
Design an adaptation solution (Managing System) using the Probe and Effector interface functions provided by the \textit{RDMSim} software as follows:\\

\subsubsection*{\textbf{A. Loading Configuration Settings and Instantiation of Probe and Effector} }
The first step in implementing the custom adaptation logic is to load the configuration settings for the experiment from the \textit{configuration.json} file and instantiation of the Probe and Effector components. The Probe and Effector will enable the communication between our custom adaptation logic and \textit{RDMSim}. This can be done by using the NetworkManagement class in your program as follows:

\begin{verbatim}
NetworkManagment network_management;
network_management=new NetworkManagment();
Probe probe;
Effector effector;
probe=network_management.getProbe();
effector=network_management.getEffector();
\end{verbatim}
 
The NetworkManagement class is responsible for loading the configuration parameters and instantiating the Probe and Effector instances.
The configuration settings include the parameters like number of simulation time steps, the number of mirrors for the RDM, number of active links and the uncertainty scenario to be considered for the experiments. The details of the configuration parameters is provided in the \textit{RDMSim Artefact:User Guide} document provided as part of \textit{RDMSimExemplar} repository. \\

\subsubsection*{\textbf{B. Monitoring of the RDMSim network using Probe functions}}
In order to monitor the RDMSim, we can use the probe functions provided in Table \ref{tab2}.
For example, to get the values of all the monitorable metrics, at a particular simulation time step, we can use the \textit{getMonitorables()} function as follows:
\begin{verbatim}
   Monitorables m=probe.getMonitorables(); 
\end{verbatim}

\subsubsection*{\textbf{C. Performing Adaptations on the RDMSim using Effector functions}}

In order to perform adaptations on the \textit{RDMSim}, we can use the Effector functions provided in Table \ref{tab3}. For example, to change the network topology at a particular timestep, we can use the \textit{setNetworkTopology()} function as follows:

\begin{verbatim}
effector.setNetworkTopology(10,"mst");
\end{verbatim}

The code above will set the Minimum Spanning Tree (MST) topology for the network at the simulation timestep 10.

A step by step implementation of the MAPE-K loop using steps A to C is provided in the \textit{User Guide} document.

\subsection*{\textbf{Step: 3 Design and Execute Experiments to test the Adaptation Logic}}
Once the adaptation solution is designed, an experiment should be designed to test the adaptation logic. For an experiment to be executed, the configuration parameters (provided in the configuration file) should be set to execute a particular simulation scenario. We have assigned some default values for the configuration parameters based on the expert knowledge provided in \cite{Paucar2020}.  You can change the number of simulation runs, the number of mirrors for the network, the  uncertainty scenario and the ranges for the different monitorables. The details for the configuration parameters are provided in the \textit{User Guide}.

\subsection*{Example: To demonstrate RDMSim working under Default and Detrimental Scenario  \begin{math}S_{1}\end{math}}

\begin{figure*}[h!]

\includegraphics[width=\textwidth,keepaspectratio]{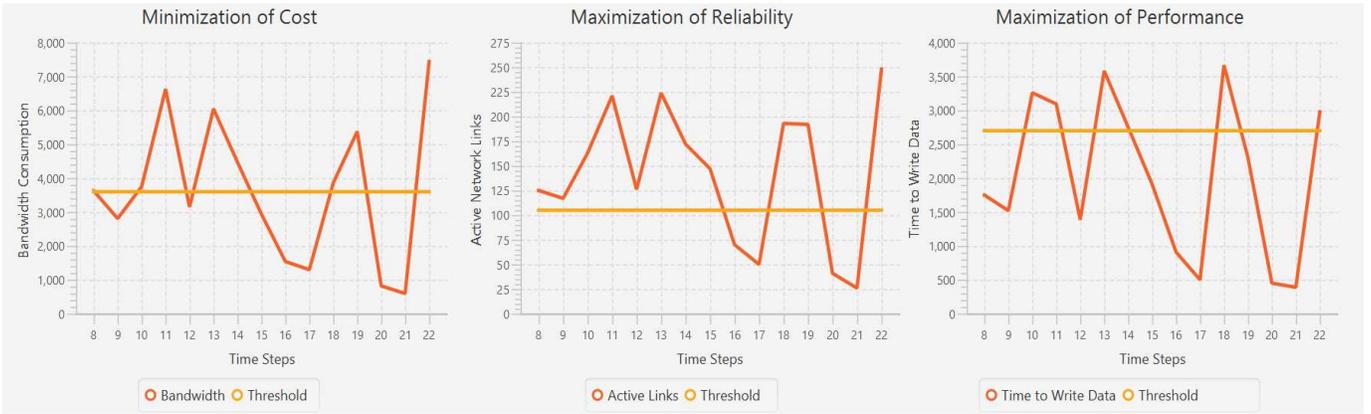}
\caption{Default Scenario}
\label{figRDMSimStableScenario}

\end{figure*}

\begin{figure*}[h!]

\includegraphics[width=\textwidth,keepaspectratio]{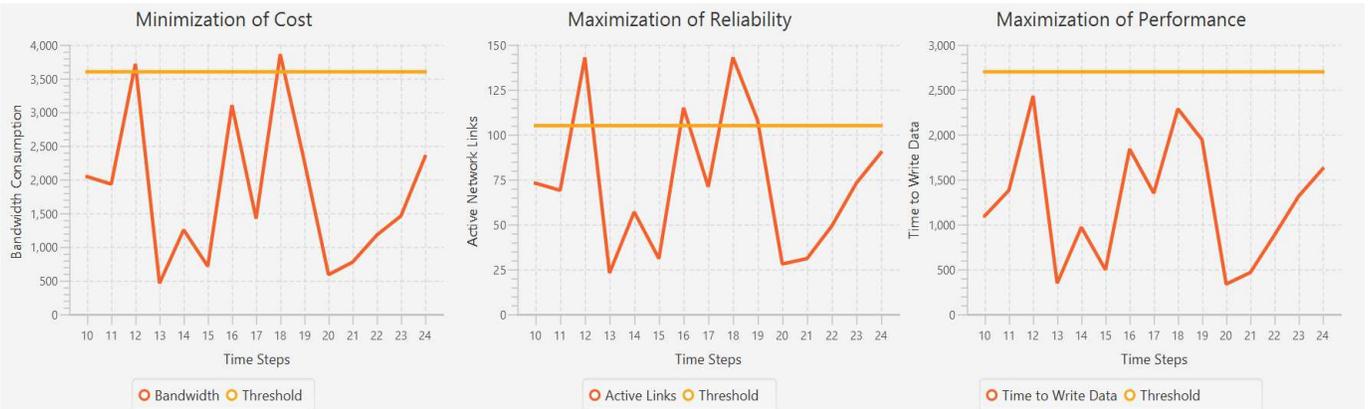}
\caption{Scenario 1}
\label{figRDMSimDetrimental}

\end{figure*}

We demonstrate the execution of experiments under both the default scenario \begin{math}S_{0}\end{math} and uncertainty Scenario \begin{math}S_{1}\end{math}. For our experiments, we consider an RDM network of 25 mirrors and 300 network links to create a fully connected network. We have set the default values for the configuration parameters in the \textit{configuration.json} file.  The satisfaction thresholds for the quality objectives have been set based on the expert knowledge provided in \cite{Paucar2020}. In order to satisfy the quality objectives of minimization of operational cost and maximization of performance, bandwidth consumption and time to write data should be minimized. Conversely, the quality objective of Maximization of Reliability requires maximization of number of active links. Based on the expert knowledge, the bandwidth consumption should be less than or equal to 40 percent to satisfy minimization of operational cost. Similarly, the time to write data should be less than or equal to 45 percent to satisfy maximization of performance.  On the other hand, number of active links should be greater than or equal to 35 percent of total links to satisfy maximization of reliability for the RDM.   Once the configuration parameters are setup, we have executed the experiments for 100 simulation runs for the scenarios as shown in Fig. \ref{figRDMSimStableScenario} and \ref{figRDMSimDetrimental}. Under default scenario, the \textit{RDMSim} will meet the satisfaction thresholds in terms of the value ranges of bandwidth consumption, active links and time to write data. Under uncertainty scenario \begin{math}S_{1}\end{math}, the different disturbance levels are introduced to reduce the number of active links affecting the reliability of the system when MST is the selected topology as shown in Fig. \ref{figRDMSimDetrimental}.

For further validation purposes, we have applied reinforcement learning based decision-making to the \textit{RDMSim}. We provide our initial evaluation results for the \textit{RDMSim} using MR-POMDP++ \cite{samin2020priority,samin2021priority} as part of the \textit{RDMSimExemplar} repository. MR-POMDP++ is based on Multi-Reward Partially Observable Markov Decision Process (MR-POMDP). MR-POMDP is a multi-objective reinforcement learning technique that considers the decision-making agent performing in a partially observable environment. MR-POMDP++ performs adaptations on the basis of the multi-objective utility value computed at each simulation time step. We have executed experiments considering a network of 25 RDM mirrors using the default configuration setup provided in the \textit{configuration} file. In order to test our decision-making techniques DeSIRE~\cite{RossBencomoSEAMS2018} and MR-POMDP++\cite{SaminSubmittedSoSyM2021}, we have also used the exemplar~\cite{IftikharRBW017}. Both exemplars ~\cite{IftikharRBW017} and \textit{RDMSim}, focus on different domains and aspects, the IoT domain and the RDM and effect on quality objectives respectively, and complement each other. 

\textbf{Discussion:} An RDM can be seen as a specific example of a more generic type of applications, where the decision making guides self-reconfiguration by identifying a target system configuration to provide the desired system behavior~\cite{ZhangChengICE2005,Goldsby2008}. A set of reconfiguration instructions to reach the desired target configuration is applied (i.e. E in MAPE). These reconfiguration instructions define an adaptation path. Several adaptation paths may be chosen, and most self-reconfiguration approaches select adaptation paths based on tradeo-offs between several objectives goals, such as performance and reliability~\cite{Goldsby2008}. As such the \textit{RDMSim} can be used to test decision-making techniques applicable to other domains as well.


\section{Conclusion}
In this paper, we have presented the \textit{RDMSim} exemplar  to provide a simulating environment for the RDM.  
\textit{RDMSim} facilitates the researchers to execute experiments in the domain of RDM. 
To the best of our knowledge, \textit{RDMSim} is the first simulator to be implemented for this domain. 
Using \textit{RDMSim}, researchers can compare their self-adaptive decision-making solutions with other techniques, including ours\cite{samin2021priority}. We have executed experiments for each scenario presented here, using our own decision-making technique, called MR-POMDP++ \cite{samin2021priority}. The results are provided in the \textit{RDMSimExemplar} repository, ready to be used for comparison purposes. Furthermore, \textit{RDMSim} also provides opportunities for researchers to design their own scenarios for experiments by modification of values in the configuration file using the instructions provided in \textit{RDMSim} user guide.
We hope that the research community will use the \textit{RDMSim} to evaluate and compare novel solutions in the area of  self-adaptive decision-making. 


\section*{Acknowledgment}
This work has been partially supported by The Lerverhulme Trust Fellowship "QuantUn: quantification of uncertainty using Bayesian Surprises" (Grant No. RF-2019-548/9) and the EPSRC Research Project Twenty20Insight (Grant No. EP/T017627/1).

\bibliographystyle{IEEEtran} 
\bibliography{References}

\end{document}